\documentclass[twocolumn,showpacs,english,prl,preprintnumbers,amsmath,amssymb,floatfix]{revtex4}

\usepackage{subfiles}

\usepackage[T1]{fontenc}
\usepackage[latin9]{inputenc}
\usepackage{dcolumn}
\usepackage{bm}
\usepackage{graphicx}
\usepackage{color}
\usepackage{esint}
\usepackage{babel}
\usepackage{amsmath,amssymb,amsthm,mathrsfs,amsfonts,dsfont}
\usepackage{slashed}
\usepackage{enumerate}
\usepackage{simplewick}

\begin{document}

\title{Spectral correlations in Anderson insulating wires }

\author{M.~Marinho and  T.~Micklitz}
\affiliation{
Centro Brasileiro de Pesquisas F\'isicas, Rua Xavier Sigaud 150, 22290-180, Rio de Janeiro, Brazil 
}

\date{\today}

\begin{abstract}

We calculate the spectral level-level correlation function of Anderson insulating wires for all three Wigner-Dyson classes. 
A measurement of its Fourier transform, the spectral form factor, is within reach of state-of-the-art 
cold atom quantum quench experiments, and  
we find good agreement  
with recent numerical simulations of the latter.
Our derivation builds on a representation of the level-level correlation function in terms of a local generating function which  
may prove useful in other contexts.  

\end{abstract}

\maketitle

{\it Introduction:---}Localization due to quantum interference in disordered systems~\cite{Anderson}
is one of the cornerstones of condensed matter physics, with 
exciting recent developments such as topological Anderson insulators~\cite{topAI3,topAI2,topAI2a,topAI2b,topAI} and 
 many-body localization~\cite{MBL1,MBL2}. Notwithstanding our profound understanding of the 
 single-particle localization problem, 
 examples of dynamical correlation functions within the Anderson insulating phase, 
 which are accessible to direct experimental verification, are rare. 
The experimental challenge is to provide set-ups 
which allow for the controlled observation of strong localization 
via some tunable parameter~\cite{StrongAL}. 
On theoretical side one faces the notorious difficulty that 
Anderson insulators reside in the
non-perturbative strong coupling limit 
of a nonlinear field theory~\cite{EfetovBook}. 

A series of recent papers proposes the direct observation of spectral correlations in 
Anderson insulators 
within a cold atom quantum quench experiment~\cite{fwd1,fwd2,fwd3,fwd4}.  
A specifically promising variant of this proposal builds on a cold atom realization of the kicked rotor
and is within reach of state-of-the-art experiments~\cite{fwd5}.
The quench protocol is summarized as follows:  
(i) A cloud of cold atoms is prepared in an initial state with a well-defined momentum
$\bold{k}_{\rm i}$, (ii) it is let to propagate freely 
under the influence of a disorder potential for some time $t$, at which
(iii) the disorder is turned off and the atomic momentum distribution 
$\rho(\bold{k}_{\rm f},t)$ is measured. 
A forward scattering peak at $\bold{k}_{\rm f}\simeq \bold{k}_{\rm i}$ is predicted to appear 
as a manifestation of an  
accumulation of those quantum coherence processes 
leading to strong Anderson localization. 
Within the quench set-up `time' plays the role of the control parameter and
the genesis of the forward scattering peak is described by the spectral form factor. 
The latter is the
Fourier transform of the connected level-level correlation function, 
\begin{align}
\label{2point}
K(\omega,L)
&=
\nu_0^{-2}
\langle \nu(\epsilon+\tfrac{\omega}{2})\nu(\epsilon-\tfrac{\omega}{2})\rangle_{\rm c},
\end{align}
where $\nu_0$ is the density of states (per spin) at the energy shell. 

The theoretical study of level-level correlations~\eqref{2point} 
in disordered systems has a 
long history~\cite{Gor'kovEliashberg,Efetov82,EfetovAdv,AltshulerShlovskii1,AltshulerShlovskii2,SivanImry,Zharekeshev,Izrailev,Shlovskii,Evangelou}. 
Analytical results are, however, only known in specific limits.  
In low-dimensional systems ($d<3$) Eq.~\eqref{2point}  
describes how Wigner-Dyson statistics at small system sizes 
$L$ evolves into Poisson 
statistics with increasing size.  
The former is associated with non-integrable chaotic dynamics, 
while the latter signals the breaking of ergodicity due to quantum localization~\cite{Porter,Haake}.
Fully uncorrelated Poisson statistics 
only realizes in the thermodynamic  
limit of unbounded system sizes, and spectral correlations 
remain 
 in finite size systems. 
It is these correlations which are accessible in the cold atom quench experiment, however,  
only asymptotic results are known for the experimentally relevant orthogonal and 
the symplectic symmetry class.

In this paper we derive the spectral level-level correlation function for Anderson insulating wires belonging 
to the three Wigner-Dyson classes. 
We show that the latter is readily calculated from the ground-state wave-function of the transfermatrix 
Hamiltonian for the supersymmetric $\sigma$-model 
reported in Ref.~\cite{Khalaf}. 
Our results are in perfect agreement 
with recent 
numerical simulations of the quench experiment~\cite{fwd5}, and 
their experimental verification 
would mark an important 
benchmark 
for our understanding of strong Anderson localization.

{\it Field theory:---}We start out from the field theory description 
of the level-level correlation function for a $d$-dimensional disordered 
system~\cite{EfetovBook,EfetovAdv,efetovreview}, 
\begin{align}
\label{KsM}
K(\omega) 
&=
{1\over 64} {\rm Re}
\langle 
\big[ \int (dx)\, 
{\rm str}\left(
k\Lambda
Q_\Lambda
\right)
 \big]^2
 \rangle_S.
\end{align} 
Here the average, 
$\langle ... \rangle_S\equiv \int {\cal D}Q(...)\exp(S)$,
is with respect to the 
diffusive nonlinear 
$\sigma$-model action, 
\begin{align}
\label{action}
S
&=  -{\pi\tilde{\nu}_0 \over 8} \int (dx)\, 
{\rm str}\left(
D(\partial_\bold{x} Q)^2 
+
2i\omega \Lambda Q
\right),
\end{align}
with $D$ the classical diffusion constant, 
`${\rm str}$' the generalization of the matrix trace to `super'-space, and $\int (dx)=1$.
The system belongs to one of the three Wigner-Dyson symmetry classes, 
characterized by the absence of  time-reversal symmetry, ${\cal T}=0$ (unitary class),
presence of time-reversal and spin-rotational symmetry, ${\cal T}^2=1$ (orthogonal class),
or 
presence of time-reversal and absence of spin-rotational symmetry, ${\cal T}^2=-1$ (symplectic class). 
Throughout the paper, we adopt the notation of Ref.~\onlinecite{EfetovBook} where
$\tilde{\nu}_0=\nu_0$ in the unitary 
and orthogonal, and  
$\tilde{\nu}_0=2\nu_0$ 
in the symplectic class. 
$Q$ is a supermatrix 
acting on an $8$-dimensional 
graded space, which is the product of two-dimensional 
subspaces, referred to as `bosonic-fermionic'  (${\rm bf}$), 
`retarded-advanced' $({\rm ra})$ and 
 `time-reversal' $({\rm tr})$ sectors. 
 Matrices $k\equiv\sigma_3^{\rm bf}$, 
 $\Lambda\equiv\sigma_3^{\rm ra}$
break symmetry in bosonic-fermionic and advanced-retarded sectors, 
respectively. 
$\Lambda$ describes the classical, diffusive fixed point
and $Q_\Lambda\equiv Q-\Lambda$ deviations from the latter. 
Drawing on the similarity of action~\eqref{action}
to Ginzburg-Landau theories for phase-transitions,
`${\rm str}(\Lambda Q)$' corresponds to a symmetry breaking term 
relevant at large level-separations or short time scales 
$t\sim\omega^{-1}\ll\Delta_\xi^{-1} \equiv \xi^2/D$, with 
 $\xi$ the localization length.    
In this 
diffusive limit $Q\simeq \Lambda$, which allows for a controlled perturbative expansion in Goldstone modes, viz.,  
the diffusion modes of the disordered single-particle system. 
Strong Anderson localization sets in at $t\sim\omega^{-1}\sim\Delta^{-1}_\xi$ when   
large fluctuations restore the symmetry in the `${\rm ra}$'-sector. 
This requires integration over the entire $Q$-field manifold and calls for 
non-perturbative methods. 
Such methods are available for quasi one-dimensional geometries $L\gg L_\perp$,
where the functional integral with action~\eqref{action} 
takes the form of a path-integral of a quantum mechanical particle with coordinate $Q$ and 
mass $\sim 1/D$, moving in a potential $\sim{\rm str}(\Lambda Q)$. 
The latter can be mapped onto the corresponding 
Schr\"odinger problem,  
and we next follow this strategy.

{\it Anderson insulating wires:---}Concentrating then on a quasi one-dimensional geometry, 
one needs to express the 
spectral correlation function Eq.~\eqref{KsM} in terms of  
eigenfunctions of the Hamiltonian for the Schr\"odinger problem, known as
the `transfermatrix Hamiltonian'. 
This has been done in previous work~\cite{EfetovBook,altlandfuchs}. 
The resulting equations for the relevant functions are, however, 
 rather complex and closed solutions for all Wigner-Dyson classes are unknown. 
 We therefore follow here 
a different route, which employs the graded symmetry of action~\eqref{action} in order  
to 
 derive Eq.~\eqref{2point} from a local generating function.
 The latter  
depends only on the ground-state wave-function, i.e. `zero-modes' of the transfermatrix Hamiltonian. 
This implies 
a significant simplification of the problem, and allows for an exact calculation of correlations~\eqref{2point}. 
We momentarily postpone the discussion of the rather technical derivation, 
and state the final expression for the generating function in case of Anderson insulating wires 
$L\gg \xi\equiv\pi\tilde{\nu}_0D/L$~\cite{fn3},
\begin{align}
\label{localCFQ0}
K(\omega)
&=
{\xi\over 2\beta L}
{\rm Re}
\int(dx)\,\partial_\eta{\cal F}(\eta,x)|_{\eta=-{i\omega\over\Delta_\xi}},
\nonumber \\
{\cal F}(\eta)
&={1\over 2}\int dQ_0\, 
{\rm str}\left(\Lambda Q_0\right)Y^2_0(Q_0).
\end{align}
Here $Y_0$ is the ground-state wave-function of the Schr\"odinger problem detailed below,  
and we introduced the symmetry parameter $\beta=1(2)$ in the 
orthogonal and symplectic (unitary) class. 
Notice that in Eq.~\eqref{localCFQ0} we already integrated out some $c$-number 
and all Grassmann variables. That is, 
$Q_0$ here depends only on $c$-number variables from the compact interval, 
$-1\leq \lambda_{\rm f} \leq 1$ (`fermionic radial variables'), 
and non-compact interval $1\leq \lambda_{\rm b}$ (`bosonic radial variables').
The precise number of radial variables depends on the symmetry class
i.e. $Q_0({\cal R})$, with 
${\cal R}=\{\lambda_{\rm f},\lambda_{\rm b}\}$,
$\{\lambda_{\rm f},\lambda_{\rm b,1},\lambda_{\rm b,2}\}$, 
and 
$\{\lambda_{\rm f,1},\lambda_{\rm f,2},\lambda_{\rm b}\}$, 
in the unitary, orthogonal and symplectic classes, respectively.
Similarly,  
$dQ_0=d{\cal R} \sqrt{g(\Lambda)}$,  
  with  
  Jacobians 
 $\sqrt{g}=1/(\lambda_{\rm b}-\lambda_{\rm f})^2$,
$\sqrt{g}=(1-\lambda^2_{\rm f})/(\lambda_1^2+\lambda_2^2+\lambda_{\rm f}^2-2\lambda_1\lambda_2\lambda_{\rm f}-1)^2$,
and 
$\sqrt{g}=(\lambda^2_{\rm b}-1)/(\lambda_1^2+\lambda_2^2+\lambda_{\rm b}^2-2\lambda_1\lambda_2\lambda_{\rm b}-1)^2$ 
for the three symmetry classes, and $d{\cal R}$ the flat measure.
For notational convenience we suppress the graded index ${\rm b,f}$ in favor of 
the `${\rm tr}$'-index $1,2$.  
The ground-state wave-function 
is a solution to the 
homogeneous equation,
\begin{align}
\label{tmeq}
\left(
-\Delta_Q + \tfrac{\eta}{2}  {\rm str}(\Lambda Q_0)
\right)
Y_0(Q_0)
&=0,
\end{align}
obeying the boundary condition 
$Y_0(\Lambda)=1$ and we recall that at 
$Q_0=\Lambda$ all radial coordinates $\lambda=1$. 
Here we introduced 
$\eta=-i\omega/\Delta_\xi$, and 
$\Delta_Q
=
{1\over \sqrt{g}}\partial_\lambda \sqrt{g} g^{\lambda\rho} \partial_{\rho}$ 
is the Beltrami-Laplace operator on the $Q_0$-field manifold 
with repeated indices running over radial variables, 
$\lambda, \rho  \in {\cal R}$
and 
 metric tensor $g^{\lambda \rho}=|\lambda^2-1|\delta^{\lambda\rho}$ in all symmetry classes, with 
$\delta^{\lambda\rho}$ the Kronecker-delta.

{\it Correlations from zero-mode:--}We may then use the Schr\"odinger equation 
to express
${\rm str}(\Lambda Q_0)\, Y_0
=-
\left(2
\Delta_Q +\eta \, {\rm str}(\Lambda Q_0)
\right)
Y'_0
$
and 
$\eta\, {\rm str}(\Lambda Q_0)\, Y_0
=-
2\Delta_QY_0$, and
arrive at
\begin{align}
\label{localCFQ1}
{\cal F}(\eta)
&= 
\int dQ_0\, 
\left(
Y'_0 
\Delta_Q
Y_0 
-
Y_0 
\Delta_Q
Y'_0 
\right),
\end{align}
where we introduced $Y_0'\equiv \partial_\eta Y_0$.
Upon partial integration this results in 
the boundary contribution
\begin{align}
\label{bgt}
{\cal F}(\eta)
&=
\int d{\cal R} \,
\partial_\lambda
 \sqrt{g} g^{\lambda\rho}
 \left( 
Y_0' 
\partial_\rho  Y_0
-
Y_0 
\partial_\rho  Y'_0
 \right).
\end{align}
At this point we notice that  
metric elements $g^{\lambda\rho}$ vanish at any 
boundary point $\lambda=1$. At the same time, the Jacobian is singular 
at $Q_0=\Lambda$ where all $\lambda_{\rm b},\lambda_{\rm f}=1$. 
To deal with this situation we regularize the integral Eq.~\eqref{bgt} 
in any of the variables $\lambda$, shifting the bound of integration to
$1^\pm\equiv1\pm\epsilon$ with $\pm$ for a bosonic/fermionic variable.
In the limit $\epsilon \searrow 0$ 
the boundary contribution 
$(\sqrt{g} g^{\lambda\rho})|_{\lambda=1^\pm}$ 
then reduces (up to a numerical factor) 
to a $\delta$-function 
 in the remaining radial coordinates, fixing $Q_0=\Lambda$. 
Noting further that 
$Y_0(\Lambda)=1$ and $Y'_0(\Lambda)=0$, we arrive at 
the remarkably simple expression
\begin{align} 
\label{localCFsimp}
{\cal F}(\eta)
&=  
-2\partial_{\lambda_{\rm f}}
Y'_0|_{Q=\Lambda},
\end{align}
where in the symplectic class $\lambda_{\rm f}$ can be either $\lambda_{1,2}$.
It can be verified that 
$\partial_{\lambda}Y'_0|_{Q=\Lambda}=\pm \partial_{\rho}Y'_0|_{Q=\Lambda}$,
where the positive sign applies if $\lambda$ and $\rho$ are both bosonic or fermionic
radial variables and the negative sign else. This guarantees that 
Eq.~\eqref{localCFsimp} does not depend on the regularization scheme, 
and
one may, e.g., symmetrize the result in the radial variables~\cite{SuppMat}.  
An equivalent relation
between the ground-state wave-function and the generating function for 
spectral correlations has been previously encountered
for the unitary class~\cite{TMLevel-Level,crr}. 
In this case the derivation is built upon a mapping of the localization  
problem in the unitary class to the three-dimensional Coulomb-problem~\cite{Skvortsov}. 
The above Eq.~\eqref{localCFsimp} shows 
that the simple relation is not accidental but applies to all Wigner-Dyson symmetry classes. 

Using then the recent results of Ref.~~\cite{Khalaf,SuppMat}
for the ground-state wave-functions 
we find
($z_\eta\equiv4\sqrt{\eta}$)
\begin{align}
\label{resFU}
{\cal F}^{\rm U}(\eta)
&= 
-8I_0(z_\eta)K_0(z_\eta),
\\
\label{resFO}
{\cal F}^{\rm O}(\eta) 
&=  
-4\left( 
I_0(z_\eta)K_0(z_\eta)
+
I_1(z_\eta)K_1(z_\eta)
 \right),
 \\
 \label{resFSp}
{\cal F}_\pm^{\rm Sp}(\eta) 
&=
-4\left( 
\left[ 
I_0(z_\eta)
\pm
1
\right]
K_0(z_\eta)
+
I_1(z_\eta)K_1(z_\eta)
 \right),
 \end{align}
where 
`${\rm U}$',
 `${\rm O}$',
 and
 `${\rm Sp}$'
refers to the unitary, orthogonal and symplectic 
 symmetry class, respectively, and
 `$+/-$' indicates an even/odd number of channels.

{\it Spectral correlations:---}From Eqs.~\eqref{resFU}-\eqref{resFSp} we find
the level-level correlations in Anderson insulating wires for all 
three Wigner-Dyson classes,
\begin{align}
\label{Kfin}
K(\omega)=
{32 \xi \over \beta L }
 {\rm Re} \,
 {\cal K}(z_\eta)|_{z_\eta=4\sqrt{-i\omega/\Delta_\xi}},
 \end{align}
where
\begin{align}
\label{resKU}
 {\cal K}^{\rm U}(z_\eta)
&=   
\left[
K_1(z_\eta) I_0(z_\eta)
-
K_0(z_\eta) I_1(z_\eta)
\right]/z_\eta,
\\
\label{resKO}
{\cal K}^{\rm O}(z_\eta)
&=  
K_1(z_\eta) I_1(z_\eta)/z^2_\eta,
 \\
 \label{resKSp}
{\cal K}_\pm^{\rm Sp}(z_\eta)
&=
 K_1(z_\eta)
 \left[
 I_1(z_\eta)
  \pm
z_\eta/2
 \right]/z^2_\eta.
 \end{align}
Eqs.~\eqref{Kfin}-\eqref{resKSp} are the main result of this paper.
Strict Poisson statistics only applies for 
$\lim_{L\to\infty}K(L,\omega)=0$, and 
correlations between localized eigenstates remain in any finite system. 
At large level-separation ($s\equiv \omega/\Delta_\xi\gg1$) 
these reflect the classically diffusive dynamics on short time-scales, on which 
quantum interference processes remain largely undeveloped. 
Correlations of close-by levels ($s\equiv\omega/\Delta_\xi\lesssim1$), 
on the other hand, store information on the long-time limit, 
i.e. the deep quantum regime 
in which remaining dynamical processes are due to 
tunneling between almost degenerate,  
far-distant localized states~\cite{Mott,log,Ivanov2012}.
The crossover between these two limits, described
by Eqs.~\eqref{Kfin}-\eqref{resKSp}, is shown in Fig~\ref{fig1}. 
For a comparison with fully chaotic systems
we also show in the inset the corresponding Wigner-Dyson correlations   
with their characteristic level repulsion 
at small 
level-separations,   
and contrasting the residual logarithmic level repulsion between 
localized states.

\begin{figure}[tt]
\begin{center}
\includegraphics[width=8.6cm]{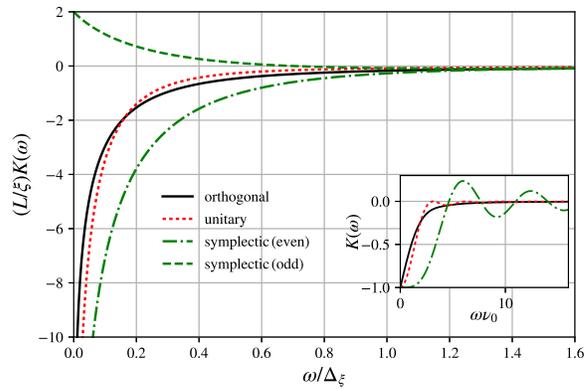}
\end{center}
\vspace{-15pt}
\caption{Level-level correlations in Anderson insulating wires 
for the Wigner-Dyson classes. The residual level-attraction  
in the symplectic class with an odd number of channels
reflects the presence of a topologically protected metallic channel.
The inset shows for comparison Wigner-Dyson spectral correlations 
of fully chaotic systems.
}
\label{fig1}
\end{figure}

From the above expression,  
one readily recovers asymptotic correlations 
of far-distant levels $s\gg1$, 
applying to all Wigner-Dyson 
classes~\cite{Gradsteyn}, 
\begin{align}
\label{lwasympt}
{L\over\xi}
K(s)
&= 
-
{1\over 4\sqrt{2}\beta }
\left(
{1\over s^{3/2}}
-
{(-1)^\beta 3\over 128s^{5/2}}
+ ...
\right), 
\end{align}
with the leading Altshuler-Shklovskii 
contribution~\cite{AltshulerShlovskii1,AkkermansMontambaux}. 
For small level-separations 
and systems 
 in the unitary, orthogonal or symplectic 
class with an even number of channels ($s \ll 1$) 
\begin{align}
\label{swasympt}
{L\over\xi}
K(s)
&= 
-
a_\beta
\left(
\log\left( 1/4s \right)
-
2\gamma+
b_\beta
+
c_\beta \pi s
+
...
\right), 
\end{align}
where 
 $\gamma\simeq0.577$ the Euler-Mascheroni constant, 
 $a_{\rm U,Sp+}=8$, $a_{\rm O}=4$,
$b_{\rm U}=0$, $b_{\rm O}=1/2$, $b_{\rm Sp+}=3/4$,
and
$c_{\rm U}=3$, $c_{\rm O}=2$, $c_{\rm Sp+}=3/2$.
In the symplectic class with an odd number of channels ($s\ll1$), 
\begin{align}
{L\over\xi}
K^{\rm Sp}_-(s)
&= 
2 - 4\pi s - 
((8s)^2/3)
\log(s)
+..., 
\end{align}
which signals the presence of a single  
topologically protected metallic channel, see also Fig~\ref{fig1}.

{\it Forward peak:---}The form factor deriving from the above results 
describes the
genesis of the forward scattering peak in the quantum quench set-up  
discussed in the introduction~\cite{SuppMat,fn6},  
\begin{align}
\label{peakU}
{\cal C}^{\rm U}_{\rm fs}(t)
&=
\theta(t)
I_0\left( 8/t \Delta_\xi  \right)
e^{-8/t \Delta_\xi },
\\
\label{peakO}
{\cal C}^{\rm O}_{\rm fs}(t)
&=
\theta(t)
\left[ 
I_0\left( 8/t\Delta_\xi  \right)
+
I_1\left( 8/t\Delta_\xi  \right)
\right]
e^{-8/t\Delta_\xi },
\\
\label{peakSp}
{\cal C}^{\rm Sp,\pm}_{\rm fs}(t)
&=
\tfrac{1}{2}
\left[
{\cal C}^{\rm O}_{\rm fs}(t)
\pm
\theta(t)
e^{-4/t\Delta_\xi }
\right].
\end{align}
Here we have normalized the peak with respect to its saturation value 
$\lim_{t\to\infty}{\cal C}_{\rm fs}(t)$. 
The forward peak in the unitary class has been calculated previously~\cite{fwd4,fn7}. 
Corresponding results for the experimentally relevant orthogonal class~\cite{Josse} 
and the symplectic class have been unknown. 
Fig.~\ref{fig2} displays a comparison of our results with recent numerical simulations of the 
quantum quench experiment in the orthogonal class~\cite{fn7}.
The solid line is Eq.~\eqref{peakO} 
and shows perfect agreement with the numerical data  without using any fitting-parameter. 
The forward peak for all Wigner-Dyson classes are displayed in the inset of Fig.~\ref{fig2}. 
${\cal C}^{\rm O}_{\rm fs}$ is readily understood as 
a sum of diagrams involving only ladders (`diffuson modes') 
 ${\cal C}^{\rm U}_{\rm fs}$, 
and diagrams containing crossed ladders (`Cooperon modes').
${\cal C}^{\rm Sp,\pm}_{\rm fs}$ follows the signal 
of the unitary class
at short times, $\tau\equiv t \Delta_\xi \ll 1$, staying a factor two below the signal in the orthogonal class, 
and becomes sensitive to the channel number once $\tau\equiv t \Delta_\xi \gtrsim 0.1$. 
For an odd channel number the signal in the symplectic class then 
 decays to zero as 
${\cal C}^{\rm Sp,-}_{\rm fs}\sim
4/\tau^2- 64/(3\tau^3)+...$, 
indicating delocalization due to the presence of the topologically protected channel.
Long and short time signals in the remaining cases 
can be summarized as 
 ($\tau\equiv t \Delta_\xi $)
\begin{align}
{\cal C}_{\rm fs}(\tau)
&=
\begin{cases}
a_\alpha \tau^{1/2}
+b_\alpha \tau^{3/2}+\dots, 	
\quad 
& s \ll 1, 
\\
1 - c_\alpha/\tau + d_\alpha/\tau^2
+\dots, 
\quad  
&  s\gg 1, 
\end{cases}
\end{align}
where 
$a_{\rm O}=2a_{\rm U, Sp+}=1/(2\sqrt{\pi})$, 
$b_{\rm O}=-2b_{\rm U}=2b_{\rm Sp+}=-1/(128\sqrt{\pi})$
$c_{\rm O,Sp+}=c_{\rm U}/2=4$, 
and $d_{\rm O}=d_{\rm U}/3=4d_{\rm Sp_+}/3$.

\begin{figure}[tt]
\begin{center}
\includegraphics[width=8.6cm]{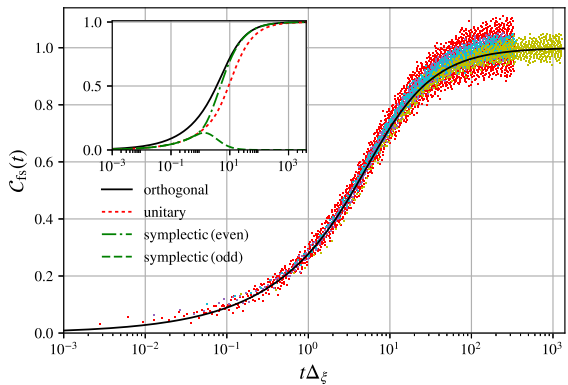}
\end{center}
\vspace{-15pt}
\caption{Forward-scattering peak in the orthogonal class. 
Points are numerical data 
 from a recent simulation of the quantum quench experiment in 
 a kicked rotor set-up~\cite{fwd5}.
  Different colors correspond to different sets of system parameters, 
  and the solid line shows Eq.~\eqref{peakO} 
 without any fitting-parameter~\cite{fnfig}.
  Insets: Forward-scattering peak for all Wigner-Dyson classes, see
  main text for discussion.
  }
\label{fig2}
\end{figure}

{\it Local generating function:---}The analysis above relied on the representation of the level-level 
correlation function in terms of a local generating function. The latter derives from the graded symmetry 
of action~\eqref{action}, 
which is evident in the 
polar parametrization $Q=UQ_0U^{-1}$~\cite{EfetovBook}.
Here matrices $U$ are diagonal in `${\rm ra}$'-sector and contain all anti-commuting variables,
while 
 $Q_0=
\cos\hat{\theta}\sigma_3^{\rm ra}-\sin\hat{\theta}\sigma_2^{\rm ra}$ 
has off-diagonal structure in the latter~\cite{SuppMat}. 
The block-diagonal matrices in  `${\rm bf}$'-sector
$\hat{\theta}={\rm diag}(i\hat{\theta}_{\rm b}, \hat{\theta}_{\rm f})_{\rm bf}$,
with $\hat\theta_{\rm b,f}$ matrices in `${\rm tr}$'-sector, are conveniently 
parametrized by 
the non-compact and compact radial variables introduced earlier,
$-1\leq \lambda_{\rm f}\equiv \cos\theta_{\rm f} \leq 1$, 
$1\leq \lambda_{\rm b}\equiv\cosh\theta_{\rm b}$~\cite{EfetovBook}. 
The graded symmetry manifests itself in the invariance of action~\eqref{action} under 
constant rotations $\bar{U}$ sharing the symmetries of 
$U$, $U\mapsto \bar{U}U$.   
This invariance can be used to linearly shift Grassmann variables in the pre-exponential correlation 
function, and  e.g. implies
 that finite contributions to the superintegral Eq.~\eqref{KsM} may only derive from
the maximal polynomial of Grassmann variables $P_{\cal G}$~\cite{Zirnbauer}. 
It is then convenient to introduce  (unnormalized)
maximal polynomials of Grassmann variables 
in retarded/advanced sectors
$P_{\cal G}^{\rm r/a}$ with 
$P_{\cal G}=P_{\cal G}^{\rm r} 
P_{\cal G}^{\rm a}$ 
and the generating function
${\cal F}(\eta,\bold{x})
 \equiv
\left\langle 
\tfrac{
\left[
{\rm str}\left(k\Lambda Q_\Lambda(\bold{x})\right)
\right]^2}
{{\rm str}\left(\Lambda Q(\bold{x})\right)
}
\right\rangle_S$. 
Notice that in the quantum-dot limit $Q$ becomes 
$\bold{x}$-independent, and the generation of 
Eq.~\eqref{KsM} by 
$\partial_\eta{\cal F}$ 
 is immediately evident. 
For general $d$-dimensional systems, on the other hand, 
a straightforward calculation shows that 
${\cal F}(\eta,\bold{x})
=
\langle 
{\rm str}( 
\cos\hat{\theta}_{\bold{x}})
P_{{\cal G},\bold{x}} 
\rangle_S$~\cite{fn10,SuppMat},
and similarly one finds
\begin{align}
&\partial_\eta{\cal F}(\eta,\bold{x})
\propto
\int (dy)\,
\langle
{\cal C}_{\bold{x},\bold{y}}
P_{{\cal G},\bold{x}}
\rangle_S,
\\
&K(\omega)
\propto
\int (dx)\int (dy)\,
\langle
{\cal C}_{\bold{x},\bold{y}}
P_{{\cal G},\bold{x}}^{\rm a}
P_{{\cal G},\bold{y}}^{\rm r}
\rangle_S,
\end{align}
with 
${\cal C}_{\bold{x},\bold{y}}
=
{\rm str}(\cos\hat{\theta}_{\bold{x}})
{\rm str}(\cos\hat{\theta}_{\bold{y}})$.
The graded symmetry can now be used to 
shift 
$P_{{\cal G},\bold{x}(\bold{y})}^{\rm r}
\mapsto
P_{{\cal G},\bold{x}}^{\rm r}
+
P_{{\cal G},\bold{y}}^{\rm r}$, 
in the first (second) term, which implies  
that Eq.~\eqref{KsM} 
is generated from the local correlation function 
for general $d$-dimensional systems. 
Indeed, keeping numerical factors one finds 
$K(\omega)=
-(16\pi\tilde{\nu}_0)^{-1}  
 {\rm Im}\, \int (dx)\, 
 \partial_\omega {\cal F}(\omega,\bold{x})$, 
 and 
 ${\cal F}(\eta)
=
{8\over\beta} 
\langle
{\rm str}(\cos\hat{\theta}_{\bold{x}})
P^0_{{\cal G},\bold{x}}\rangle_S$ with $P^0_{\cal G}$ now 
the normalized maximal polynomial of Grassmann variables~\cite{SuppMat}. 
Upon integration over the latter 
 one arrives at 
Eq.~\eqref{localCFQ0} 
for the Anderson insulating wires. 
Notice that similar ideas 
have  previously 
been applied in the context of the
replicated $\sigma$-model~\cite{Smith} and  
parametric correlations~\cite{altshulerlg}. The representation of level-level correlations 
in terms of the local generating function  
may also prove useful in other contexts~\cite{future}.

{\it Summary:---}We have shown that spectral correlations in the Wigner-Dyson classes 
can be calculated within the supersymmetric $\sigma$-model from a local generating function.
In Anderson insulating wires  this reveals
a simple relation between level-level correlations and 
the ground-state wave-function 
of the transfermatrix Hamiltonian, 
which allowed us to derive spectral correlation functions 
for all Wigner-Dyson classes.  
The experimental observation of the spectral form factor   
is within reach of state-of-the-art 
cold atom quantum quench experiments, and a parameter-free comparison 
of our findings with recent numerical simulations of the latter shows perfect agreement. 
The experimental verification of the results reported here 
would mark an important 
benchmark  
for our understanding of strong Anderson localization.

{\it Acknowledgements:---}We thank G.~Lemari\'e for providing us with their simulation 
data of the quantum quench experiment. 
T.~M.~acknowledges financial support by Brazilian agencies CNPq and FAPERJ.

\clearpage

\title{Supplemental Material to:
``Spectral correlations in Anderson insulating wires" 
}

\author{M.~Marinho and  T.~Micklitz}

\affiliation{
Centro Brasileiro de Pesquisas F\'isicas, Rua Xavier Sigaud 150, 22290-180, Rio de Janeiro, Brazil 
}

\date{\today}


\begin{abstract}

In this Supplemental Material, we summarize polar coordinates for $Q$-matrices 
and zero-modes of the transfermatrix Hamiltonian in the three Wigner-Dyson classes. 
We present details on the local generating function, the evaluation 
of the boundary integrals, and the calculation of the forward scattering peak.

\end{abstract}

\maketitle

\section{Polar coordinates and zero modes}

For convenience of the reader we  
summarize polar coordinates and ground-state wave-functions (`zero-modes') 
for the three Wigner-Dyson classes. 

{\it Polar coordinates:---}The decomposition 
$Q=UQ_0U^{-1}$,
was introduced in the main text.
Here 
\begin{align}
&Q_0=\begin{pmatrix}
\cos\hat{\theta} & i\sin\hat{\theta} \\
-i\sin\hat{\theta}  & -\cos\hat{\theta} 
\end{pmatrix}_{\rm ra},
\quad
U=\begin{pmatrix}
u & 0 \\
0 & v
\end{pmatrix}_{\rm ra},
\end{align}
with 
$\hat{\theta}
={\rm diag}(i\hat{\theta}_{\rm bb}, \hat{\theta}_{\rm ff})_{\rm bf}$, 
and matrices $\hat{\theta}_{\rm bb,ff}$ in `${\rm tr}$'-sector 
depending on the fundamental symmetries,
\begin{align}
\hat{\theta}_{\rm bb}
&=
\begin{pmatrix}
\theta_{\rm b} & 0 \\
0 & \theta_{\rm b}
\end{pmatrix}_{\rm tr},
\quad
\hat{\theta}_{\rm ff}
=\begin{pmatrix}
\theta_{\rm f} & 0 \\
0 & \theta_{\rm f}
\end{pmatrix}_{\rm tr},
\quad ({\rm U})
\\
\hat{\theta}_{\rm bb}
&=
\begin{pmatrix}
\theta_{{\rm b},1} & \theta_{{\rm b},2} \\
\theta_{{\rm b},2} & \theta_{{\rm b},1}
\end{pmatrix}_{\rm tr},
\quad
\hat{\theta}_{\rm ff}
=\begin{pmatrix}
\theta_{\rm f} & 0 \\
0 & \theta_{\rm f}
\end{pmatrix}_{\rm tr},
\quad ({\rm O})
\\
\hat{\theta}_{\rm bb}
&=
\begin{pmatrix}
\theta_{\rm b} & 0 \\
0 & \theta_{\rm b}
\end{pmatrix}_{\rm tr},
\quad
\hat{\theta}_{\rm ff}
=\begin{pmatrix}
\theta_{{\rm f},1} & \theta_{{\rm f},2} \\
\theta_{{\rm f},2} & \theta_{{\rm f},1}
\end{pmatrix}_{\rm tr},
\quad ({\rm Sp}),
\end{align}
where compact and  
non-compact parameters $0<\theta_{\rm f} <\pi$ 
and
$\theta_{\rm b}>0$, respectively~\cite{SMEfetovBook}. 
In the main text we use radial variables 
$-1\leq \lambda_{\rm f}\equiv \cos\theta_{\rm f}\leq 1$, 
$1\leq \lambda_{\rm b}\equiv \cosh\theta_{\rm b}$, and suppress 
the graded index ${\rm b}$, ${\rm f}$ in favor of the `${\rm tr}$'-index, $1,2$.
It is convenient to parametrize 
$U=U_1U_2$,
with 
$U_i
={\rm diag}(u_i,  v_i)_{\rm ra}$ and 
$u_1$, $v_1$ containing all Grassmann variables and 
$u_2$, $v_2$ depending only on $c$-numbers. 
The latter are of no further 
relevance for us, and the former can be parametrized as  
$u_1(\hat{\eta})
= 1 - 2\hat{\eta}+ 2\hat{\eta}^2 - 4\hat{\eta}^3+ 6 \hat{\eta}^4$
and 
$v_1(\hat{\kappa})
= 1 - 2\hat{\kappa}+ 2\hat{\kappa}^2 - 4\hat{\kappa}^3+ 6 \hat{\kappa}^4$, 
where 
\begin{align}
\hat{\eta}&=
\begin{pmatrix}
0 & \eta^\dagger \\
-\eta & 0
\end{pmatrix}_{\rm bf},
\quad 
\hat{\kappa}
=
\begin{pmatrix}
0 & i\kappa^\dagger \\
-i\kappa & 0
\end{pmatrix}_{\rm bf},
\end{align}
with
\begin{align}
\eta
&=
\begin{pmatrix}
\eta_1 & 0\\
0 & -\eta_1^*
\end{pmatrix}_{\rm tr},
\quad
\kappa
=
\begin{pmatrix}
\kappa_1 & 0\\
0 & -\kappa_1^*
\end{pmatrix}_{\rm tr},
\quad ({\rm U})
\\
\eta
&=
\begin{pmatrix}
\eta_1 & \eta_2\\
-\eta_2^* & -\eta_1^*
\end{pmatrix}_{\rm tr},
\quad
\kappa
=
\begin{pmatrix}
\kappa_1 & \kappa_2 \\
-\kappa_2^* & -\kappa_1^*
\end{pmatrix}_{\rm tr},
\quad ({\rm O, Sp}).
\end{align}
We recall that
$(\eta_i^*)^*=-\eta_i$ and similar for $\kappa_i$, 
and notice that
$[u_1(\hat{\eta})]^{-1}=u_1(-\hat{\eta})$, 
and 
$[v_1(\hat{\kappa})]^{-1}=v_1(-\hat{\kappa})$~\cite{SMEfetovBook}.

{\it Zero-modes for Wigner-Dyson classes:---}The ground-state 
wave-functions  for the transfermatrix Hamiltonian 
of the Wigner-Dyson symmetry classes,  
recently derived in Ref.~\onlinecite{SMKhalaf}, 
read ($p=\sqrt{(\lambda+1)/2}$) 
\begin{widetext}
\begin{align}
&Y^{\rm U}_0(\lambda_{\rm b},\lambda_{\rm f})
=
4\sqrt{\eta}
\Big(
p_{\rm f} 
K_0(4\sqrt{\eta} p_{\rm b})I_1(4\sqrt{\eta}p_{\rm f})
 + 
p_{\rm b}
K_1(4\sqrt{\eta}p_{\rm b})I_0(4\sqrt{\eta}p_{\rm f})
\Big),
\\
&Y^{\rm O}_0(\lambda_1,\lambda_2,\lambda_{\rm f})
=
\sqrt{\eta}
\Big(
4p_1p_2
I_0(4\sqrt{\eta}p_{\rm f}) K_1(4\sqrt{\eta}p_1p_2)
+
{1\over p_{\rm f}}
( 1 + \lambda_1 +\lambda_2 +\lambda_{\rm f} )
I_1(4\sqrt{\eta}p_{\rm f}) K_0(4\sqrt{\eta}p_1p_2)
\Big),
\\
&Y^{\rm Sp}_0(\lambda_{\rm b},\lambda_1,\lambda_2)
=
\sqrt{\eta}
\Big(
4p_1p_2
I_1(4\sqrt{\eta}p_1p_2) K_0(4\sqrt{\eta}p_{\rm b})
+
{1\over p_{\rm b}}
( 1 + \lambda_1 +\lambda_2 +\lambda_{\rm b} )
I_0(4\sqrt{\eta}p_1p_2) K_1(4\sqrt{\eta}p_{\rm b})
\Big),
\\
&Y^{\rm Sp}_\pm(\lambda_{\rm b},\lambda_1,\lambda_2)
=
Y^{\rm Sp}_0(\lambda_{\rm b},\lambda_1,\lambda_2)
\pm 
Y^{\rm Sp}_0(\lambda_{\rm b},-\lambda_1,-\lambda_2).
\end{align} 
\end{widetext}

\section{Generating function}

Starting out from the generating function, 
introduced in the main text,
 \begin{align}
{\cal F}(\eta,\bold{x})
 &=
\left\langle 
{
\left[
{\rm str}\left(k\Lambda Q_\Lambda(\bold{x})\right)
\right]^2
\over
{\rm str}\left(\Lambda Q(\bold{x})\right)
}
\right\rangle_S,
\end{align}
it is verified that in polar coordinates
\begin{align}
{\rm str}\left(\Lambda Q\right)
&=
2\,{\rm str}( 
\cos\hat{\theta}), 
\\
{\rm str}\left(k\Lambda Q\right)
&=
{\rm str}\left(k \Lambda U_2^{-1} U^2_1U_2 Q_0\right),
\end{align}
where in the second line 
we employed cyclic invariance of the trace and anti-commutation of 
$\hat{\eta}$ and $k$.
 $k U_2^{-1}U_1^{2}U_2$ is a diagonal matrix in `${\rm ra}$'-sector,  
with `${\rm rr}$'-block 
\begin{align}
&
[ k U_2^{-1}U_1^{2}U_2]_{\rm rr}
\nonumber\\
&=
u_2^{-1}
k
\left[
1
-
8
\left(
\begin{smallmatrix}
\eta^\dagger\eta
-
4 (\eta^\dagger \eta)^2
& 0
\\ 0 
&
\eta \eta^\dagger 
-
4 (\eta  \eta^\dagger)^2 
\end{smallmatrix}\right)_{\rm bf}
\right]
u_2
+
... .
\end{align}
The omitted terms `$...$' summarize contributions  
off-diagonal in `${\rm bf}$'-sector, i.e. vanishing under the trace.
A similar expression involving $\kappa$ holds for the `${\rm aa}$'-block.

In the unitary class 
 $\eta$ is diagonal in the 
 `${\rm tr}$'-sector, i.e. 
$\eta^\dagger\eta=-\eta \eta^\dagger
=\eta_1^\ast\eta_1\openone_{\rm tr}$,
and 
$(\eta^\dagger  \eta)^2=0$.
In the remaining classes $\eta$
has off-diagonal structure in the `{\rm tr}'-sector and 
$(\eta^\dagger  \eta)^2
=-(\eta  \eta^\dagger )^2
=2\eta_1^\ast \eta_2^\ast \eta_1\eta_2\openone_{\rm tr}$.
Recalling that   
in any correlation function we only need 
to account for contributions 
independent of Grassmann variables 
or being of maximal order of the latter 
(see e.g. Ref.~\onlinecite{SMZirnbauer}) 
we may simplify
\begin{align}
[ k U_2^{-1}U_1^{2}U_2]_{\rm rr}
&=
k
-
{\cal P}_{\cal G}^{\rm r}
+
...,
\end{align}
with 
${\cal P}_{\cal G}^{\rm r}\propto \eta_1^\ast\eta_1$ 
and
${\cal P}_{\cal G}^{\rm r}\propto \eta_1^\ast \eta_2^\ast \eta_1\eta_2$  
the (unnormalized) maximal polynomials 
of Grassmann variables in the retarded sector 
for the unitary, respectively, the orthogonal and symplectic
symmetry classes. 
A similar calculation for the `${\rm aa}$'-block  
then leads to 
 \begin{align}
{\rm str}\left(k\Lambda Q\right)
&=
{\rm str}( 
[2k+P_{\cal G}^{\rm r}
+
P_{\cal G}^{\rm a}]
\cos\hat{\theta}) 
+...,
\end{align}
where omitted contributions `$...$' give again  vanishing contributions to the correlation function.
That is, keeping only terms with non-vanishing contributions to the correlation function,
\begin{align}
\left[
{\rm str}\left(k\Lambda Q_\Lambda \right) 
\right]^2
&=
[{\rm str} 
(2k (\cos\hat{\theta}- \openone_4) )  
]^2
+
2 [{\rm str}( 
\cos\hat{\theta})  
]^2
P_{\cal G}, 
\end{align}
with 
$P_{\cal G}=
P_{\cal G}^{\rm r}
P_{\cal G}^{\rm a}$ 
the (unnormalized) maximal polynomial in Grassmann variables.
Similarly,
 \begin{align}
 \label{SMgf}
{\cal F}(\eta,\bold{x})
 &=
\left\langle 
{\rm str}( 
\cos\hat{\theta})
P_{\cal G} 
\right\rangle_S,
\end{align}
where we noticed that 
boundary terms 
(i.e. those independent of Grassmann variables) here vanish.
Finally, 
\begin{widetext}
\begin{align}
&\int (dx)\int (dy)\,
\langle
{\cal F}(\eta,\bold{x})
{\rm str}\left(\Lambda Q(\bold{y})\right)
\rangle_S
=
2
\int (dx)\int (dy)\,
\langle
{\rm str}(\cos\hat{\theta}(\bold{x}))\,
{\rm str}(\cos\hat{\theta}(\bold{y}))\,
P_{\cal G}(\bold{x})
\rangle_S,
\\
&\int (dx)\int (dy)\,
\langle
{\rm str}\left(k\Lambda Q_\Lambda(\bold{x})\right)
{\rm str}\left(k\Lambda Q_\Lambda(\bold{y})\right)
\rangle_S
=
2
\int (dx)\int (dy)\,
\langle
{\rm str}(\cos\hat{\theta}(\bold{x}))\,
{\rm str}(\cos\hat{\theta}(\bold{y}))\,
P_{\cal G}^{\rm a}(\bold{x})\,
P_{\cal G}^{\rm r}(\bold{y})
\rangle_S,
\end{align}
\end{widetext}
and both expression become identical upon 
 shifting 
$P_{\cal G}^{\rm r}(\bold{x})
\mapsto
P_{\cal G}^{\rm r}(\bold{x})
+
P_{\cal G}^{\rm r}(\bold{y})$, 
in the first term, 
and 
$P_{\cal G}^{\rm r}(\bold{y})
\mapsto
P_{\cal G}^{\rm r}(\bold{y})
+
P_{\cal G}^{\rm r}(\bold{x})$  in the second term, as discussed in the main text.
To fix all factors, we can then 
normalize the maximal polynomial of Grassmann
variables by considering the 
quantum-dot limit, where $Q(\bold{x})=Q_0$ is space-independent. That is,
for  
\begin{align}
P_{\cal G}
&={8\over\beta} P^0_{\cal G},
\end{align}
with $P^0_{\cal G}$ the normalized maximal polynomial of Grassmann variables 
(introduced in the main text) 
Eq.~\eqref{SMgf} recovers Wigner-Dyson statistics~\cite{SuppMatfn1}.

\section{Boundary terms}

In the main text we have derived the 
following 
representation for the generating function
for level-level correlations in one of 
the three Wigner-Dyson classes,
\begin{align}
\label{SMappbt}
{\cal F}(\eta)
&=
\int d{\cal R} \,
\partial_\lambda
 \sqrt{g} g^{\lambda\rho}
 \left( 
Y_0' 
\partial_\rho  Y_0
-
Y_0 
\partial_\rho  Y_0'
 \right),
\end{align}
where $Y_0'\equiv\partial_\eta Y_0$. 
We next discuss different ways to regularize Eq.~\eqref{SMappbt}, all  
leading to the same result. 

{\it Unitary class:---}Calculations are the simplest in the unitary class, 
where Eq.~\eqref{SMappbt} takes the form
\begin{widetext}
\begin{align}
{\cal F}^{\rm U}(\eta)
&=
\int_1^\infty d\lambda_{\rm b}
\int_{-1}^{1} d\lambda_{\rm f}\,
\left(
\partial_{\lambda_{\rm f}}  
{(1-\lambda_{\rm f}^2)G_{\rm f}(\lambda_{\rm b},\lambda_{\rm f})\over (\lambda_{\rm b}-\lambda_{\rm f})^2}
+
\partial_{\lambda_{\rm b}} 
{(\lambda_{\rm b}^2-1) G_{\rm b}(\lambda_{\rm b},\lambda_{\rm f}) \over (\lambda_{\rm b}-\lambda_{\rm f})^2}
\right),
\end{align}
\end{widetext}
and to simplify notation we introduced 
$G_{\rm f,b}
\equiv
Y_0' 
\partial_{\lambda_{\rm f,b}}   Y_0
-
Y_0 
\partial_{\lambda_{\rm f,b}}   Y_0'$. 
As discussed in the main text, the first (second) term vanishes 
at the boundary $\lambda_{\rm f}=1$ 
($\lambda_{\rm b}=1$). At the same time the Jacobian 
$\sqrt{g}=1/(\lambda_{\rm b}-\lambda_{\rm f})^2$
diverges at $\lambda_{\rm f}=\lambda_{\rm b}=1$.
To deal with this situation we regularize (some of) the radial variables shifting the 
limit of integration $1\mapsto 1^\pm\equiv 1\pm \epsilon$,
where the positive/negative sign applies for the bosonic/fermionic radial variable. 
Regularizing e.g. in the bosonic radial variable,
\begin{align}
{\cal F}^{\rm U}(\eta)
&=
2\lim_{\epsilon\to0}
\int_0^\infty 
dx_{\rm f}
{\epsilon G_{\rm f}(1, 1)
\over 
(x_{\rm f} +\epsilon)^2}
=
2G_{\rm f}(1, 1),
\end{align}
where $x_{\rm f}=1-\lambda_{\rm f}$
and we  used that in the limit $\epsilon\searrow 0$
we only 
need to keep leading contributions in $x_{\rm f}\ll1$. 
Similarly,  one finds upon
regularization in the fermionic radial variable
${\cal F}^{\rm U}(\eta)=-2G_{\rm b}(1, 1)$.
If, on the other hand, both variables are regularized,
\begin{align}
{\cal F}^{\rm U}(\eta)
&=
2\lim_{\epsilon\to0}
\left(
\int_\epsilon^\infty 
dx_{\rm f}
{\epsilon G_{\rm f}(1, 1)
\over 
(x_{\rm f} +\epsilon)^2}
-
\int_\epsilon^\infty 
dx_{\rm f}
{\epsilon G_{\rm f}(1, 1)
\over 
(x_{\rm f} +\epsilon)^2}
\right)
\nonumber\\
&
=
G_{\rm f}(1, 1)
-
G_{\rm b}(1, 1),
\end{align}
and the same is found 
in polar coordinates upon excluding $r=0$, 
i.e.
${\cal F}^{\rm U}(\eta)=
2\lim_{\epsilon\to0}
\int_0^{\pi/2} d\phi\,
\left(
{\cos^2\phi \, G_{\rm f}(1,1)\over(\cos\phi+\sin\phi)^2}
-
{\sin^2\phi \, G_{\rm b}(1,1)\over(\cos\phi+\sin\phi)^2}
\right)$.
Noting that
$G_{\rm f}(1,1)=-G_{\rm b}(1,1)$ 
it is verified that all of the above procedures give the same result.

{\it Other classes:---}Calculations in the orthogonal and symplectic classes are more involved but follow 
the same line. Upon regularization, the boundary contribution $(\sqrt{g}g^{\lambda\rho})|_{\lambda=1^\pm}$
in Eq.~\eqref{SMappbt} reduces (up to a numerical factor) to a $\delta$-function in the remaining 
radial coordinates, fixing the latter to $\lambda=1$.
Zero-modes in all symmetry classes share the properties that   
$Y_0(\Lambda)=1$, $Y_0'(\Lambda)=0$,  and
$\partial_{\lambda}Y'_0|_{Q=\Lambda}=\pm \partial_{\rho}Y'_0|_{Q=\Lambda}$.
Here the positive sign applies if $\lambda$ and $\rho$ are both bosonic or fermionic
radial variables and the negative sign else, and we recall that $\Lambda$ corresponds  
to setting all radial variables $\lambda=1$.  These 
properties guarantee that the result is independent of which/how many integration limits are shifted.  

Changing e.g. in the orthogonal class 
the integration limit in the fermionic variable,
\begin{widetext}
\begin{align}
{\cal F}^{\rm O}(\eta)
&=
\lim_{\epsilon\to0}
\int_1^\infty d\lambda_1
\int_1^\infty d\lambda_2\,
\int_{-1}^{1-\epsilon} d\lambda_{\rm f}\,
\partial_{\lambda_{\rm f}} \sqrt{g} g^{\lambda_{\rm f} \lambda_{\rm f}} 
G_{\rm f}(\lambda_{\rm f},\lambda_1,\lambda_2)
\nonumber\\
&=
\lim_{\epsilon\to0}
\int_1^\infty d\lambda_1
\int_1^\infty d\lambda_2
\,
{4\epsilon^2 G_{\rm f}(1,\lambda_1,\lambda_2)
\over \left( 
\epsilon^2
+
\lambda_1^2+\lambda_2^2 
-
2\lambda_1\lambda_2
-
2\epsilon(1-\lambda_1\lambda_2)
\right)^2
}
=
2G_{\rm f}(1,1,1),
\end{align}
\end{widetext}
while a similar calculation with any of the two integration limits $\lambda_{1,2}$ shifted 
gives
${\cal F}^{\rm O}(\eta)
=-2G_{\rm b}(1,1,1)$, etc.
As stated in the main text, we may, therefore, equivalently present the final results
in a symmetrized form, e.g.
\begin{align}
{\cal F}^{\rm U}_0(\eta)
&=
\left(
\partial_{\lambda_{\rm b}} 
-
\partial_{\lambda_{\rm f}}
 \right)
Y_0'|_{\lambda_{\rm b}=\lambda_{\rm f}=1},
 \end{align}
in the unitary class, and similarly for the other symmetry classes.

\section{Forward peak}  

The time evolution of the forward scattering peak is, up to a normalization factor, 
given by the Fourier transform of the level-level correlation function. 
E.g. 
in the unitary class ($\eta=-i\omega/\Delta_\xi$)
\begin{align}
{\cal C}_{\rm fs}(t)
&
\propto
 \Delta_\xi^{-1} \int_{-\infty}^\infty d\omega\, e^{-i\omega t}
\partial_\eta K_0(4\sqrt{\eta})  I_0(4\sqrt{\eta})
\nonumber \\
&=
i\Delta_\xi t \theta(t)
\int_{0}^\infty dz
\, e^{-z t\Delta_\xi}
\Big(
K_0\left(4i\sqrt{z}\right)  
I_0\left(4i\sqrt{z}\right)
\nonumber\\
&\qquad \qquad 
-
K_0\left(-4i\sqrt{z}\right)  
I_0\left(-4i\sqrt{z}\right)
\Big)
\nonumber \\
&=
\Delta_\xi t  \theta(t)
\int_{0}^\infty dz
\, e^{-z t\Delta_\xi}
J_0\left(4\sqrt{z}\right)  
J_0\left(4\sqrt{z}\right)
\nonumber \\
&=
\theta(t)
 I_0\left( {8\over \Delta_\xi t} \right)
e^{-{8\over \Delta_\xi t}},
\end{align}
where in the third line we used that 
$K_n(-z)
=(-1)^nK_n(z)+\left( \log(z) - \log(-z) \right) I_n(z)$, 
$I_n(-z)
=(-1)^nI_n(z)$, and $I_n(iz)=i^nJ_n(z)$. 
A similar calculation for the remaining classes, 
leads to Eqs.~(19)-(21) stated in the main text.



\begin{thebibliography}{99}
\bibitem{fn3}
 The localization length  here is always that for the orthogonal class, i.e.
 $\xi_{\rm O}=\xi$ while $\xi_{\rm U}=2\xi$ and $\xi_{\rm Sp}=4\xi$.
\bibitem{fn6}
That is,  up to a normalization factor
${\cal C}_{\rm fs}(t)= \int_{-\infty}^\infty d\omega\, e^{-i\omega t}
\partial_\eta {\cal F}(\eta)$.
\bibitem{fn7}
Notice that here we use $\Delta_\xi\equiv\Delta^{\rm O}_\xi=D/\xi_{\rm O}^2$ 
with $\xi_{\rm O}$ the localization length in the orthogonal class, 
and e.g. $\Delta^{\rm U}_\xi=\Delta^{\rm O}_\xi/4$, see also Ref.~\onlinecite{fn3}.
\bibitem{fnfig}
The universal curve is obtained after  
accounting for finite Ehrenfest-times and with Heisenberg-times $\sim1/\Delta_\xi$ 
independently 
determined from the wave-packet dynamics (see Ref.~\onlinecite{fwd5} for further details).
\end{thebibliography}
\begin{thebibliography}{99}
\bibitem{SuppMatfn1}
Notice that we here used that integrals over 
variables parametrizing $U_2$ are normalized.
\end{thebibliography}
\end{document}